\documentclass[aps,prb,longbibliography,superscriptaddress,twocolumn,showkeys,amsmath,amssymb,floatfix]{revtex4-2}

\usepackage[english]{babel}
\usepackage{amsmath,amssymb,bbm,mathrsfs,bm,braket,color,graphicx,comment,amsfonts,dsfont}
\usepackage{siunitx,physics}
\usepackage[colorlinks,citecolor=blue,urlcolor=blue]{hyperref}

\begin{document}

\title{Anisotropic optics and gravitational lensing of tilted Weyl fermions}

\author{Viktor K\"{o}nye}
\affiliation{Institute for Theoretical Solid State Physics, IFW Dresden and W\"{u}rzburg-Dresden Cluster of Excellence ct.qmat, Helmholtzstr. 20, 01069 Dresden, Germany}
\author{Lotte Mertens}
\affiliation{Institute for Theoretical Solid State Physics, IFW Dresden and W\"{u}rzburg-Dresden Cluster of Excellence ct.qmat, Helmholtzstr. 20, 01069 Dresden, Germany}
\affiliation{Institute for Theoretical Physics, University of Amsterdam, Science Park 904, 1098 XH Amsterdam, The Netherlands}
\author{Corentin Morice}
\affiliation{Laboratoire de Physique des Solides, CNRS UMR 8502, Université Paris-Saclay, F-91405
Orsay Cedex, France}
\author{Dmitry Chernyavsky}
\affiliation{Institute for Theoretical Solid State Physics, IFW Dresden and W\"{u}rzburg-Dresden Cluster of Excellence ct.qmat, Helmholtzstr. 20, 01069 Dresden, Germany}
\author{Ali~G.~Moghaddam}
\affiliation{Department of Physics, Institute for Advanced Studies in Basic Sciences (IASBS), Zanjan 45137-66731, Iran}
\affiliation{Computational Physics Laboratory, Physics Unit, Faculty of Engineering and
Natural Sciences, Tampere University, FI-33014 Tampere, Finland}
\author{Jasper van Wezel}
\affiliation{Institute for Theoretical Physics, University of Amsterdam, Science Park 904, 1098 XH Amsterdam, The Netherlands}
\author{Jeroen van den Brink}
\affiliation{Institute for Theoretical Solid State Physics, IFW Dresden and W\"{u}rzburg-Dresden Cluster of Excellence ct.qmat, Helmholtzstr. 20, 01069 Dresden, Germany}
\affiliation{Institute for Theoretical Physics, TU Dresden, 01069 Dresden, Germany
}

\date{\today}

\begin{abstract}
We show that tilted Weyl semimetals with a spatially varying tilt of the Weyl cones provide a platform for studying analogues to problems in anisotropic optics as well as curved spacetime. Considering particular tilting profiles, we numerically evaluate the time evolution of electronic wave packets and their current densities. We demonstrate that electron trajectories in such systems can be obtained from Fermat's principle in the presence of an inhomogeneous, anisotropic effective refractive index. On the other hand, we show how the electrons dynamics reveal gravitational features and use them to simulate gravitational lensing around a synthetic black hole. 
These results bridge optical and gravitational analogies in Weyl semimetals, suggesting novel pathways for experimental solid-state electron optics.
\end{abstract}

\maketitle

Inspiration drawn from light optics has driven important fundamental and applied breakthroughs in electronic condensed matter systems.
It led to the manipulation of electrons using inhomogeneous potentials \cite{vanHouten1995}, negative refractive indices and Veselago lensing \cite{Cheianov2007,Cserti2007}, tailoring the effective mass of electrons \cite{Engheta2012}, and a geodesic description for random variations in surface height \cite{Dahlhaus2010}. Even more exotic phenomena such as electron beam collimation \cite{Louie2008supercollimation} and cloaking \cite{Liao2012Cloaking,Liao2013,Fleury2013,pendry2006controlling} have been proposed. 
In the past decade, some of these advanced theoretical ideas have been realized in experiments which brought about a new era in the field of \emph{solid state electron optics} \cite{Gossard2001coherent,lee2015observation}.
In this context, we consider systems with a specific spatially-varying hopping profile rather than the more traditional inhomogeneous potential landscape \cite{vanHouten1995},  which allows us to devise an electronic version of anisotropic optics \cite{Born1999,Rivera1995,Rudzki2001}.
In the long wavelength limit this inhomogeneous hopping profile maps to a system with Weyl cones whose tilting depends on position. 

This is of particular interest as close analogies can be drawn between electronic Weyl cones and the light cones defined in general relativity \cite{Volovik:2016kid, volovik2003universe,kedem2020black}.
Recent developments showed behaviours analogous to gravitational horizons for a range of phenomena in inhomogeneous systems related to Weyl semimetals, including eternal slowdown and Unruh radiation \cite{guan2017artificial, huang2018black, Ojanen2019, Morice2021, morice2021scipostcore, moghaddam2021engineering, Sabsovich2021, konye2022horizon, mertens2022thermalization}.
There are already several proposals for experimentally realizing the modulations in tilting as a function of space, including the use of structural distortions, spin textures, and external position-dependent driving \cite{Vozmediano2018,Ghimire2019,Ray2022,Bardarson2019,Weststrom2017,long2020,lee2016}.
These novel possibilities enrich the field of \emph{analogue gravity} that have been proposed in a range of other types of systems \cite{Barcelo2001, philbin2008, Carusotto2008, barcelo2011analogue, Boada2011, Riera2012, sindoni2012emergent, nguyen2015, Minar2015, Celi2017, Calabrese2017, duine2017, Rodriguez2017, Kosior2018, kollar2019hyperbolic, Nissinen2020, Boettcher2020, lapierre2020, Jafari2019, Boettcher2022, Mula2021, de2021artificial, Stalhammar2021,Weinfurtner2011, steinhauer2016, Hu2019}.

\begin{figure}[t]
\centering
\includegraphics[width=8.6cm]{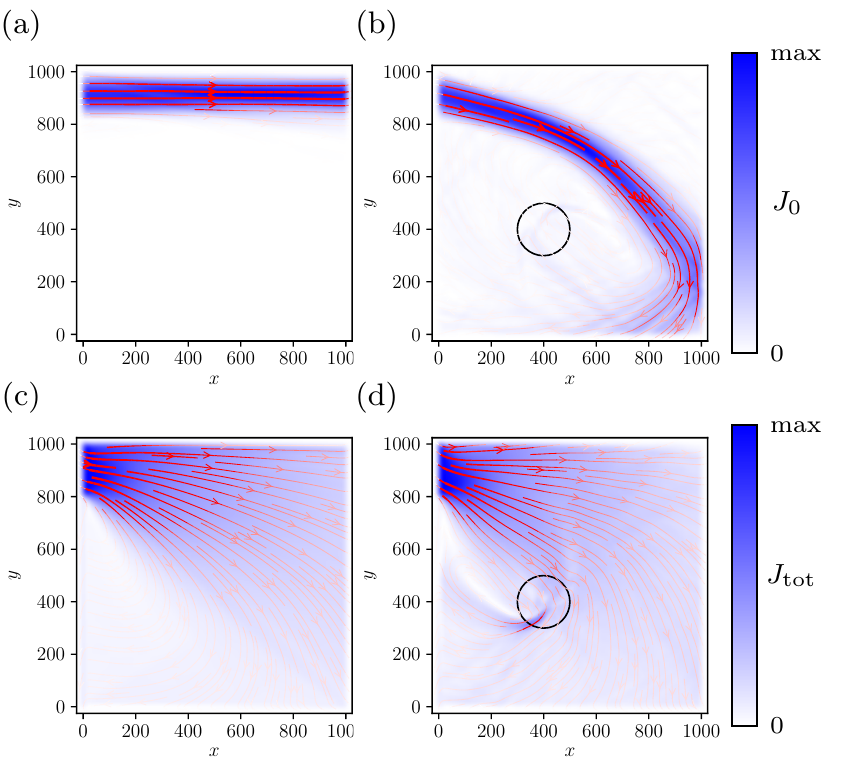}
\caption{\label{fig:current} Current density emanating from a lead in the upper left corner, flowing through a scattering region without (panels (a) and (c)) and with (panels (b) and (d)) a horizon (black circle). In (a) and (b) only a single mode is shown, while in (c) and (d) all modes are present. Color indicates the absolute value of the current density, red arrows its direction (see SM for details).}
\end{figure}

Here, we consider in particular lattice systems of tilted Weyl cones whose tilt varies slowly compared to the lattice spacing, and we numerically evaluate the quantum dynamics of wave packets in such systems. 
Intriguingly, we show that the propagation in these systems can be understood from two seemingly disconnected perspectives: (a) in analogy to anisotropic continuum optics through Fermat's principle; (b) mimicking analogue black holes and gravitational lensing for electrons.
These features are illustrated in Fig.~\ref{fig:current} where the current density is depicted both in the absence (panels (a) and (c)) and presence (panels (b) and (d)) of tilting modulations mimicking a curved spacetime with a black hole horizon (black circle). Electrons injected from a lead attached to the upper left side are seen to bend around the synthetic black hole.

Systems with position-dependent tilting of Weyl cones
provide a unique platform to study phenomena which brings anisotropic optical analogies and connections to gravitational physics together. To showcase the different facets of the approach, we address it at different levels of theory, including semiclassical trajectories, quantum wave packet dynamics, and experimentally accessible electronic current densities resulting from the presence of an event horizon.

{\it Anisotropic optics of tilted Weyl cones ---}
We consider the motion of wave packets in inhomogeneous systems with tilted Weyl cones.
At every real space position the Hamiltonian can be described as a tilted Weyl cone in the local momentum space, and we allow this tilt to change as a function of position.
This system is described by the real space Hamiltonian
\begin{equation}
\label{eq:tiltedWeyl}
H = \sum_j -i \left[ \sigma_j - t_j(\vb{r})\sigma_0 \right] \partial_j + i\frac{1}{2}\partial_j t_j(\vb{r})\sigma_0,
\end{equation}
where the tilting $\vb{t}$ depends on position, and $\sigma_j$ are the Pauli matrices.
The last term enforces the Hermiticity of the Hamiltonian.

In homogeneous systems, with constant ${\bf t}$, the Schr\"{o}dinger equation can be easily solved using plane waves, but here the Hamiltonian is spatially varying.
We assume that the tilting varies slowly with position, which allows us to use the eikonal approximation \cite{Born1999}. Details of the following derivation are available in the Supplemental Material (SM) \cite{supplemental-material}.
Starting from a plane wave ansatz with position-dependent amplitude and phase, $\Phi_\alpha = A_\alpha(\vb{r},t) e^{i \phi (\textbf{r},t)}$, and assuming slow variations of the amplitude, constant energy, and no interband mixing, we obtain the eikonal equation
\begin{equation}
E = \pm k - \sum_j t_j k_j,
\label{eq:Eikonal}
\end{equation}
where $E$ is the fixed energy, and the wave vector is given by $\vb{k}=\grad{\phi}$.

\begin{figure}[t]
\centering
\includegraphics[width=8.6cm]{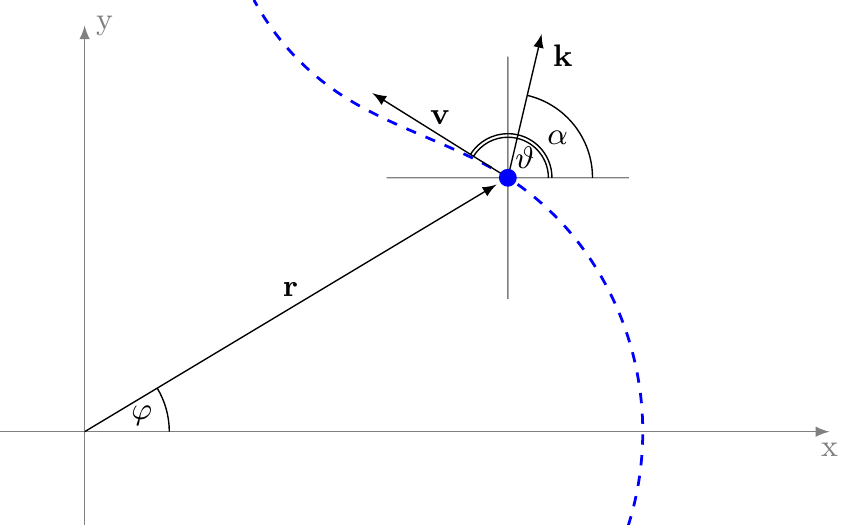}
\caption{\label{fig:fermat} Definition of the vectors of importance for the variational problem based on Fermat's principle. $\vb{v}$ is the group velocity tangential to the trajectory (blue dashed line) with angle $\vartheta$. $\vb{k}$ is the wave vector with angle $\alpha$.  $r$ and $\varphi$ are the polar coordinates.}
\end{figure}

This problem is analogous to solving the Schr\"{o}dinger equation of free electrons in an inhomogeneous potential \cite{vanHouten1995}, or solving the wave equation of light in inhomogeneous refractive media \cite{Born1999}. Fermat's principle stipulates that along the trajectories for which the variation of the optical path length is vanishing, we get constructive interference, and these will form the classical optical rays or trajectories.
They can be obtained from a variational problem as
\begin{align}
\label{eq:fermat}
   \delta \int  \vb{k}\cdot\dd{\vb{r}} = \delta \int  k\cos{(\alpha-\vartheta)}\dd{l} \equiv \delta \int  n_{\text{eff}}\dd{l}=0,
\end{align}
where $n_{\mathrm{eff}}$ is the effective refraction index and the angles are defined in Fig.~\ref{fig:fermat}.
Notice that the effective refraction index depends on the direction of the group velocity ($\vartheta$) and that the wave vector and the group velocity are not necessarily parallel ($\alpha\neq\vartheta$). These are major differences between the current problem and optics or electrons subjected to an inhomogeneous potential. They are classic signatures of anisotropic optics, seen also in for example crystal optics \cite{Born1999} or seismic wave propagation \cite{Rudzki2001}.
A well known example of its effect in classical anisotropic optics concerns refraction at perpendicular incidence to an interface, where the outgoing ray may leave at a finite angle. An analog of this phenomenon is possible in our system too, where an electron does not refract at a perpendicular angle upon impinging perpendicularly at an interface (see SM for a video of this effect).

To illustrate the rich anisotropic physics we can access in this inhomogeneously tilted Weyl system, we focus on two-dimensional systems with circular symmetry where the tilt can be expressed as
\begin{equation}
    t_i({\bf r}) = \frac{x_i}{r} t(r).
    \label{eq:tilting-profile}
\end{equation}
Transforming to polar coordinates (see Fig.~\ref{fig:fermat}), we can express the effective refraction index as \cite{supplemental-material}
\begin{align}
\label{eq:neff}
    n_{\text{eff}}   &= \frac{E}{v}, & v &= \sqrt{1-t^2\sin^2{(\vartheta-\varphi)}}-t\cos{(\vartheta-\varphi)}.
\end{align}
with $v$ the magnitude of the group velocity.
Here, the energy $E$ only appears as a multiplicative factor and does not affect the variational problem.
Therefore, particles with different energies will trace out the same trajectories in space.
This is a unique feature of the linear dispersion of the Weyl cones, and is not valid if the dispersion includes non-linear terms.

To solve the variational problem, we can take the commonly used arc-length parametrization of the trajectory \cite{Rivera1995,Rudzki2001} and find the Euler-Lagrange equations \cite{supplemental-material}.
We then get a system of differential equations that, given an initial position ($r_0, \varphi_0$) and an initial direction ($\vartheta_0$), uniquely defines a classical trajectory for the electron. Importantly, it does not give any information regarding dynamics, such as the time it takes to travel along a trajectory.\\

{\it Gravitational lensing ---}
We now turn to a specific system with rotational symmetry, in which the anisotropic physics is directly inherited from relativistic light propagation in curved spacetime. Indeed, it has been shown recently that systems featuring tilted Weyl cones can be understood as analogues to propagation in spacetime in the proximity of a black hole horizon \cite{Volovik:2016kid, volovik2003universe, Weststrom2017, guan2017artificial, huang2018black, Ojanen2019, Morice2021, morice2021scipostcore, moghaddam2021engineering, Sabsovich2021, konye2022horizon, mertens2022thermalization}.

The connection to general relativity can be understood by choosing the following circularly symmetric metric in polar coordinates \cite{gautreau1978, kraus1994simple}
\begin{equation}
\label{eq:metric_spherical}
    ds^2 = \big[t^2(r)-1\big]dT^2-2 t(r) dr dT+dr^2+r^2d\varphi^2.
\end{equation}
where $T$ denotes the temporal coordinate and $t(r)$ is the radial tilt defined in Eq.~\eqref{eq:tilting-profile}. This metric has a horizon when $t(r) = 1$.
As shown in Ref.~\cite{konye2022horizon} the massless Dirac equation in a curved background defined by this metric leads to Eq.~\eqref{eq:tiltedWeyl} in terms of tilted Weyl cones.

By choosing the specific tilting dependence
\begin{equation}
    t^2(r) = \frac{2 M}{r},
\end{equation}
the metric of Eq.~\eqref{eq:metric_spherical} describes a Schwarzschild black hole with radius $r_s = 2M$ written in the Painlev\'e-Gullstrand coordinate system \cite{volovik2021macroscopic, supplemental-material}. 

\begin{figure}[t]
    \centering
    \includegraphics[width=.8\columnwidth]{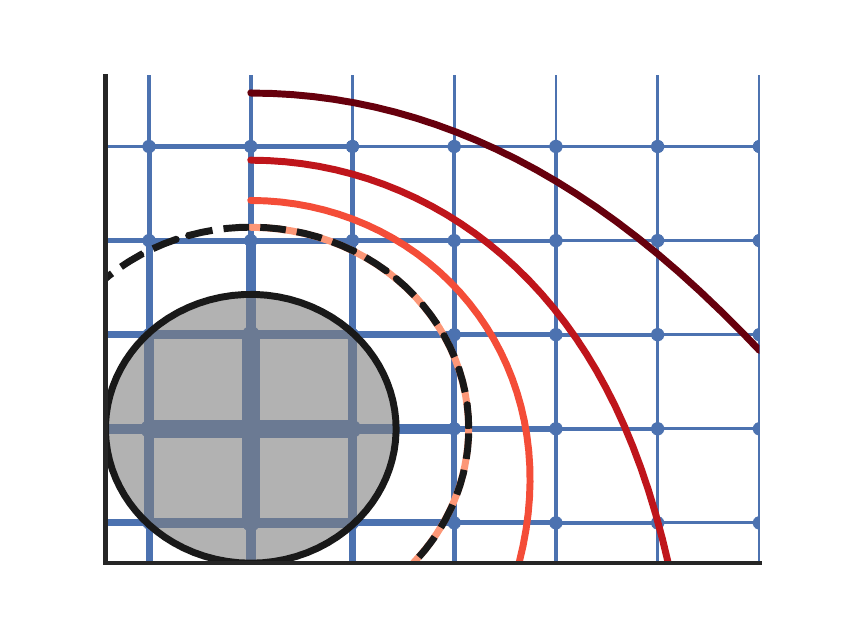}
    \caption{Geodesics of light around a black hole (grey sphere) in the Painlev\'e-Gullstrand coordinates. The red lines represent light with a different initial position and zero radial velocity. The light gets deflected inward by the gravitational potential. The dashed black line is the photon-sphere, the stable orbit of light. The blue dots and lines give a schematic representation of the lattice in a corresponding Weyl semimetal implementation, with thickness indicating the strength of hopping.}
    \label{fig:GeoSchw}
\end{figure}

We are interested in the deflection of light-like trajectories evidencing gravitational lensing, and its analogues in Weyl semimetals with inhomogeneous tilt profiles.
The light-like trajectories given by geodesics of the metric in Eq.~\eqref{eq:metric_spherical} can be numerically calculated and are shown in Fig.~\ref{fig:GeoSchw} (for more details, see the SM). The different lines represent light rays approaching the black hole horizon with different impact factors. The black hole, whose horizon is represented by the black circle, can be clearly seen to curve the light inwards.
At a distance $3r_s/2$ from the black hole centre, we find the photon sphere (dashed circle) defined by light having a circular orbit around the black hole. 

Note that the Painlev\'e-Gullstrand metric, which we use here, is related to the Schwarzschild metric by a transformation of the time coordinate while leaving the position coordinates unchanged. The spatial trajectories traced out by the light, and in particular the lensing due to the gravitational potential, are the same in both metrics, although the dynamics along those trajectories may differ (see SM for details).\\

\begin{figure*}[thb!]
\centering
\includegraphics[width=2 \columnwidth]{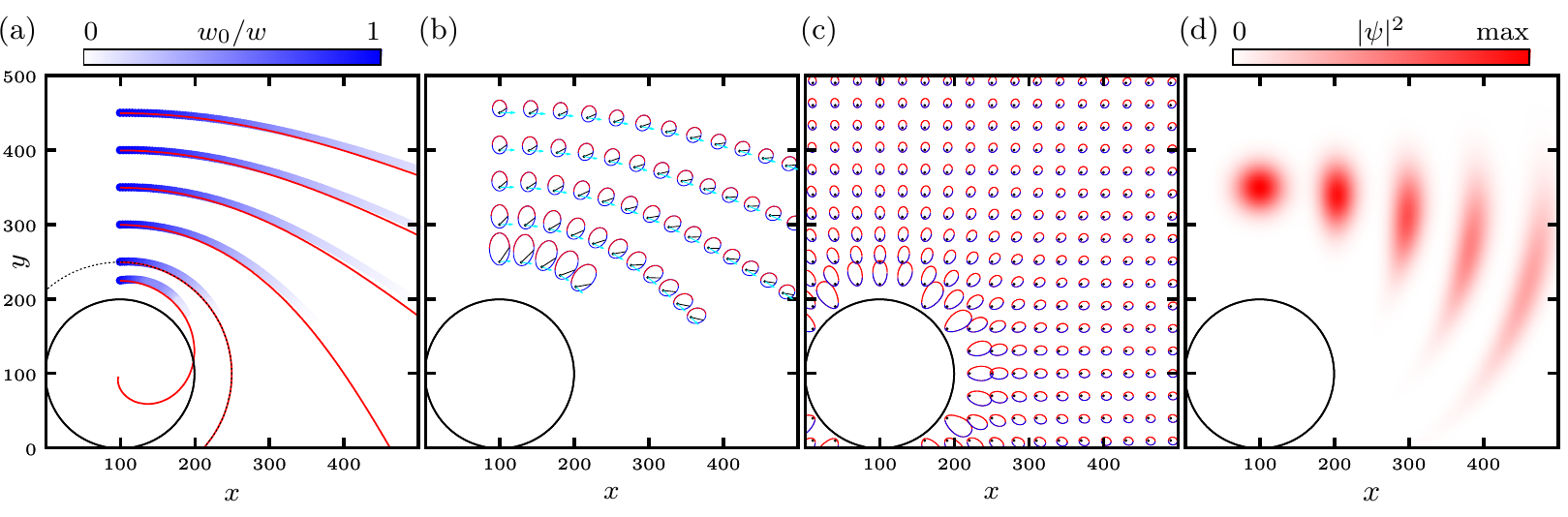}
\caption{\label{fig:lensing} (a) Calculated trajectories of wave packets using the Chebyshev expansion method (blue), and trajectories obtained using Fermat's principle (red), in an inhomogeneously tilted Weyl system. The black solid line delineates the transition between type-I and type-II Weyl cones, which corresponds to the black hole horizon in the analogous gravitational system. The dotted black line corresponds to the photon sphere. The shade of blue denotes the inverse of the width $w$ of the wave packet relative to the initial width $w_0$. (b) Constant energy lines in the local Bloch Hamiltonians along five different trajectories. The black and cyan lines indicate the wave vector and group velocity respectively. (c) Constant energy lines for local Bloch Hamiltonians throughout the whole system. In panels (b) and (c) the $\Gamma$ point of the local Brillouin zone is represented by the black point and the color of the energy lines denotes the sign of $v_y$ red (blue) is positive (negative). (d) Probability density of a propagating wave packet shown at five different time steps.}
\end{figure*}

{\it Schwarzschild black hole with tilted Weyl nodes ---} In the following we consider a two-dimensional lattice model that hosts tilted Weyl-cones at low energies, introduced in Ref. \cite{konye2022horizon}:
\begin{align}
        \label{eq:H12D}
        \nonumber
        H =& (\sigma_x - t_x\sigma_0)\sin{k_x}+(\sigma_y - t_y\sigma_0)\sin{k_y}+\\
&+ \sigma_z (2-\cos{k_x}-\cos{k_y}).
\end{align}
where $H$ is the local Bloch Hamiltonian. We focus on the specific case that maps to the Schwarzschild black hole where the radial tilting defined in Eq.~\eqref{eq:tilting-profile} is $t(r) = \sqrt{r_s/r}$.
This Hamiltonian is schematically depicted in Fig.~\ref{fig:GeoSchw}: the lattice in blue has thicker edges where the hopping proportional to $\sigma_0$ is larger. The tilting therefore increases when approaching $r=0$. The black line corresponds, in the electronic system, to the transition between type-I and type-II Weyl cones. In the gravitational system, this line corresponds to the black hole horizon.

To establish the extent of the connection between this tight-binding model and gravitational physics, we investigate the relation between the propagation of the wave packets in the tight-binding model and the geodesics of light defined by the metric. 
The wave packet trajectories in the tight-binding model for different impact parameter are shown in Fig.~\ref{fig:lensing}a.
These were obtained using two different methods. First, using Fermat's principle (Eq.~\eqref{eq:fermat}) and solving the Euler-Lagrange equation numerically (red curves in Fig.~\ref{fig:lensing}a).
Second, by simulating the full quantum wave packet dynamics using the Chebyshev expansion method \cite{Kosloff1994,Weisse2006} on the tight-binding Hamiltonian given by Eq.~\eqref{eq:H12D} (blue lines in Fig.~\ref{fig:lensing}a).
The trajectories calculated using the two methods are in close agreement with each other and with the geodesics in Fig.~\ref{fig:GeoSchw}.
All trajectories are deflected, and decreasing the impact factor increases the curvature of the trajectory. If the starting position is on the photon sphere, the trajectory becomes circular, while smaller impact factors imply collapse towards the centre of the analogue black hole.

To understand the origin of the lensing effect in the context of tilted Weyl semimetals, we show a representation of the Fermi surface defined by the local Bloch Hamiltonian, at the energy of the wave packet, along five specific trajectories in Fig.~\ref{fig:lensing}b, and on the whole lattice in Fig.~\ref{fig:lensing}c.
This energy is conserved and determines the dynamics of the wave packet.
Far away from $r=0$, the Fermi surfaces are almost circular, with the $\Gamma$ point at the center. Getting closer to the horizon, the Fermi surfaces become elliptical, with the $\Gamma$ point heavily shifted towards the horizon.
In particular, this causes the wave vector and group velocity to not be parallel, similar to Eq.~\eqref{eq:fermat}, as depicted in Fig.~\ref{fig:lensing}b.
Along the trajectory, the wave vector must change continuously while conserving the energy.
This, together with the rotation of the ellipses along the trajectory, leads to a rotation of the group velocity and to a bending of the trajectories around the black hole.

Moving beyond the semiclassical description, we can consider the actual wave packets and their broadening inside the system. 

To quantify this, we calculate the maximal full width at half maximum ($w$) of the wave packet, represented in Fig.~\ref{fig:lensing}a by the color scale.
The ratio of the initial width to the actual width ($w_0/w$) goes to zero for a delocalized wave function and is equal to one for the initial state.
The $w_0/w$ ratio decreases for all trajectories during their propagation through the system, but the decrease is much faster for trajectories that come close to the horizon.
Snapshots of the propagation of one particular wave packet are depicted in Fig.~\ref{fig:lensing}d.

The initially well-localized wave packet gets compressed in the direction of propagation and becomes wider in the perpendicular direction.
The semiclassical description using Fermat's principle breaks down once the wave packet becomes too delocalized in real space, since the point-like particle approximation is no longer valid.
Furthermore, it also breaks down once the wave packet gets too squeezed, making it less localized in momentum space, and mixing in components for which the dispersion is no longer linear.\\

{\it Current density ---}
Having characterized the wave packet trajectories and their connection to geodesics and Fermat's principle, we turn to the effect of lensing on electronic currents, which can be accessed experimentally.
We consider a multi-terminal transport setup, where we attach leads to all sides of a square scattering region and use the Kwant package \cite{Groth2014} to calculate the current density associated to the scattering states (see SM for details of the setup and calculations). The results are summarized in Fig.~\ref{fig:current}.

When inputting a single propagating mode with large momentum, shown in panels (a) and (b), the current densities resemble the semiclassical trajectories.
Indeed, without any Weyl cone tilting, the current density is rectilinear and only sizable in a horizontal region, while including inhomogeneous Weyl cone tilting causes the current density to display a clear lensing effect.

Next, to consider the total transport through the scattering region, we include all propagating modes in the lead. Summing over these, we obtain the current density profile corresponding to what would be measured in a typical transport measurement, as shown in panels (c) and (d).
The presence of the inhomogeneous Weyl cone tilting then bends the current density, which results in a significantly larger current reaching the bottom lead, although the effect is less prominent than when inputting a single mode.
In terms of multi-terminal conductances, increasing the radius of the overtilted region would make the conductance to the bottom lead grow while reducing the one to the right lead.\\

{\it Conclusion ---}
We showed that tilted Weyl semimetals with inhomogeneous tilting are a promising platform to study phenomena related to anisotropic optics.
The trajectories of electrons can be obtained using an anisotropic effective refractive index.
We discussed how this model could be related to massless particles propagating around a Schwarzschild black hole and suggested a setup to measure this phenomenon experimentally.
With the rapidly growing number of realizations of tilted Weyl semimetals both in real condensed matter systems \cite{Soluyanov2015, Sun2015, huang2018black} and meta-materials \cite{Zangeneh-Nejad2020, Li2022}, and different methods to engineer inhomogeneity \cite{Vozmediano2018, Jeroen2019rotation, Bardarson2019, Weststrom2017, long2020, lee2016} in the tilting, the effects discussed in this work can be experimentally accessed.
Using the analogy to anisotropic optics one can devise effects and corresponding experimental realizations that would otherwise be impossible in systems of electrons in an inhomogeneous potential.

{\it Note added ---}
While preparing this manuscript we became aware of a related work \cite{Tobias} that considers an alternative approach to gravitational lensing in a related system using semiclassical dynamics discussing Berry curvature effects. Where our studies overlap, we get consistent results.

\begin{acknowledgments}
We are grateful to I.C. Fulga for helpful discussions. We thank U. Nitzsche for technical assistance. 
We also thank the Würzburg-Dresden Cluster of Excellence on Complexity and Topology in Quantum Matter – ct.qmat (EXC 2147, Project No. 390858490)
\end{acknowledgments}

\bibliography{references}

\end{document}